\documentclass[%
 aip,
 amsmath,amssymb,
 reprint,%
]{revtex4-1}

\usepackage{graphicx}
\usepackage{dcolumn}
\usepackage{bm}
\usepackage{amsmath}
\usepackage{upgreek}
\usepackage{comment}
\usepackage[utf8]{inputenc}
\usepackage[T1]{fontenc}
\usepackage{mathptmx}
\usepackage{etoolbox}

\makeatletter
\def\@email#1#2{%
 \endgroup
 \patchcmd{\titleblock@produce}
  {\frontmatter@RRAPformat}
  {\frontmatter@RRAPformat{\produce@RRAP{*#1\href{mailto:#2}{#2}}}\frontmatter@RRAPformat}
  {}{}
}%
\makeatother
\begin{document}

\preprint{AIP/123-QED}

\title{Digital Etching of Single-Crystalline Silicon Fin Barriers Down to Sub-10~nm}
\author{Yu Wu }
\thanks{These authors contributed equally to this work.}
 \affiliation{Department of Electrical $\&$ Computer Engineering, University of California, Santa Barbara, CA, 93106, USA}
\author{Teun A. J. van Schijndel}
\thanks{These authors contributed equally to this work.}
 \affiliation{Department of Electrical $\&$ Computer Engineering, University of California, Santa Barbara, CA, 93106, USA}
 \author{Anthony P. McFadden}
 \affiliation{National Institute of Standards and Technology, Boulder, CO, 80305, USA}
\author{Wilson J. Y\'{a}nez-Parre\~{n}o}
\affiliation{Department of Electrical $\&$ Computer Engineering, University of California, Santa Barbara, CA, 93106, USA}%

 \author{Raymond W. Simmonds}
 \affiliation{National Institute of Standards and Technology, Boulder, CO, 80305, USA}
 
\author{Christopher J. Palmstrøm}
\email{cjpalm@ucsb.edu}
 \affiliation{Department of Electrical $\&$ Computer Engineering, University of California, Santa Barbara, CA, 93106, USA}
 \affiliation{Materials Department, UC Santa Barbara, CA}
 
\date{\today}

\begin{abstract}
This work presents a fabrication process to realize sub-10 nm-thick, high-aspect-ratio Si fins using high-resistivity float-zone single-crystal Si wafers. Following anisotropic wet etching, a digital etching process based on repeated rapid thermal annealing (RTA) oxidation and vapor-phase hydrofluoric acid (VHF) etching of the resulting oxide was developed, enabling thinning of Si fins into the sub-10~nm regime. Cross-sectional scanning transmission electron microscopy confirms fin thicknesses as thin as 6~nm, although with a tapered profile resulting from the initial anisotropic wet etch. Tapering of the fin was reduced by using an N$_2$/O$_2$ (5:1) ambient together with a reduced ramp rate during the temperature ramping for the RTA oxidation process. A thickness variation of 3~nm along a 1.2~$\upmu$m tall fin was achieved on a fin with 12~nm near the top and 15~nm near the base. The developed process provides a pathway for fabricating Josephson junctions, setting a foundation toward reproducible and scalable superconducting qubit fabrication with single-crystal Si dielectric barriers using readily available Si processing technology.
\end{abstract}

\maketitle
A superconducting transmon qubit consists of a shunt capacitor in parallel with a nonlinear inductor, a Josephson Junction (JJ), forming an anharmonic LC circuit whose two lowest energy levels serve as the qubit states \cite{koch2007charge}.  Since its introduction in 2007, the transmon qubit has enabled dramatic advances in quantum computing \cite{arute2019quantum,kjaergaard2020superconducting}. It mitigates charge noise by shunting a small-area JJ with a large planar capacitor \cite{koch2007charge}.

Recently, Bland et al. demonstrated 2D transmon qubits with a lifetime up to 1.68 milliseconds by utilizing the tantalum-on-high-resistivity Si platform, substantially reducing bulk dielectric loss \cite{bland2025millisecond}. In these devices, the suppression of substrate and surface losses was sufficient to render junction-related decoherence observable, and further improvement in times required a low-contamination junction deposition process \cite{bland2025millisecond}. This suggests that two-level system (TLS) defects in or near the JJ barrier are becoming a limiting factor for further reduction of energy loss. These TLS are commonly attributed to the structural disorder of amorphous dielectric materials, including the tunnel barrier, which permits a large number of dangling bonds and bonding configurations \cite{martinis2005decoherence, muller2019towards,faoro2006quantum,ku2005decoherence}.

One promising route towards low-loss JJs is utilizing single-crystalline dielectric materials, which lack the structural disorder believed to be responsible for TLS in amorphous oxides. Single-crystalline Al$_2$O$_3$ barriers have been demonstrated to significantly eliminate approximately 80\% of the two-level defect states compared to amorphous AlO$_x$ barriers \cite{oh2006elimination}. The residual TLS is attributed to the non-epitaxial interface between the Al$_2$O$_3$ barrier and the Al top electrode, where interfacial oxygen ions bond equally well to the Al, reproducing the bonding disorder of an amorphous AlO$_x$ barrier \cite{oh2006elimination}. Therefore, further reduction of interfacial TLS motivates the exploration of other high-quality dielectrics for single-crystalline barriers and epitaxial superconductors \cite{oh2006elimination,oh2005low}.

High-resistivity undoped float-zone Si is one of the lowest-loss dielectric materials at cryogenic temperatures and microwave frequencies \cite{checchin2022measurement}. In addition, Si is widely used in the semiconductor industry and possesses mature fabrication processes. Recently, a successful demonstration of transmons using crystallographic selective wet chemical etching of Si to create single-crystalline Si fins as parallel plate capacitor dielectrics achieved state-of-the-art performance, with low-power internal quality factors exceeding $500{,}000$ in lumped-element resonators and transmon energy relaxation $T_1$ times greater than $25~\upmu$s \cite{mcfadden2025fabrication}, confirming that both the etched Si fin surface and the $\text{Al/Si}$ interface have sufficient quality to support functional transmon circuits. These results indicate that single-crystalline Si fins are promising candidates not only as low-loss capacitor dielectrics but also as dielectric barriers in JJs.

 \begin{figure*}[t]
\includegraphics[width=0.9\textwidth]{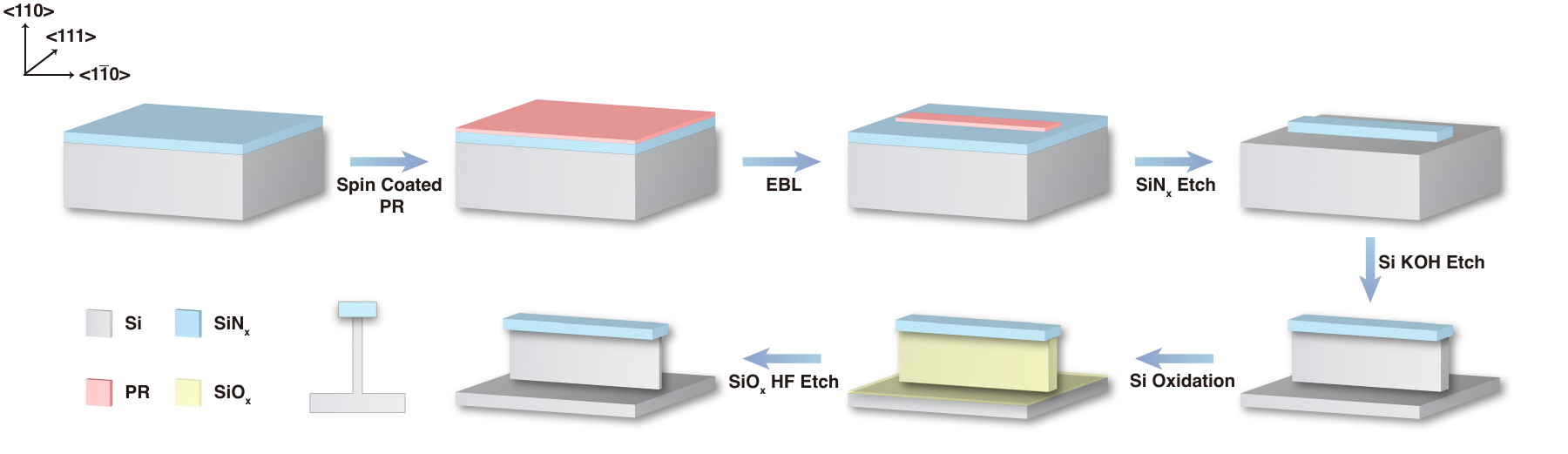}
\caption{\label{fig:1} Schematic of the fabrication workflow for Si fin fabrication including electron beam lithography (EBL) defining the hard mask pattern, inductively coupled plasma (ICP) etching the SiN$_x$ hard mask, Si etching through KOH, followed by Si digital etching (repeated Si oxidation and SiO$_x$ etching). The layers are Si (grey), SiN$_x$ (blue), photoresist (PR, red), and SiO$_x$ (yellow).}
\end{figure*}

However, unlike amphorous Al$_2$O$_3$ barriers that can be formed by oxidizing and annealing polycrystalline Al, Si single-crystal barriers would conventionally have to be grown on the superconductor surface. Although metals can be grown epitaxially on semiconductors, the differences in crystal structure, bonding, and surface energies make it very challenging to grow high-quality epitaxial semiconductors and insulators on a metal \cite{palmstrom1995epitaxy}. An alternative approach is to create a vertical single-crystalline Si fin structure and deposit a superconductor on both sides of the fin to form a superconductor/single crystal Si/superconductor structure \cite{goswami2022towards}. First-principles calculations of Schottky barrier heights combined with tunneling model calculations indicate that the Josephson current decays exponentially with the Si barrier thickness, and achieving a Josephson critical current density of 2~$\sim$~3x10$^5$~A$/$m$^2$, requires a Si barrier thickness in the 5~$\sim$~10~nm regime for Al contacts \cite{nangoi2024first}. Previous fabrication of Si fins using conventional KOH wet etching achieved a thickness of 50~nm, and digital etching was proposed as a route to the 5~$\sim$~10~nm regime but was not demonstrated \cite{goswami2022towards}. In this work, an optimized fabrication process of wet chemical etching followed by digital etching was developed, which realized Si fin thicknesses in the sub-10~nm regime.

\begin{figure*}[th]
\includegraphics[width=0.9\textwidth]{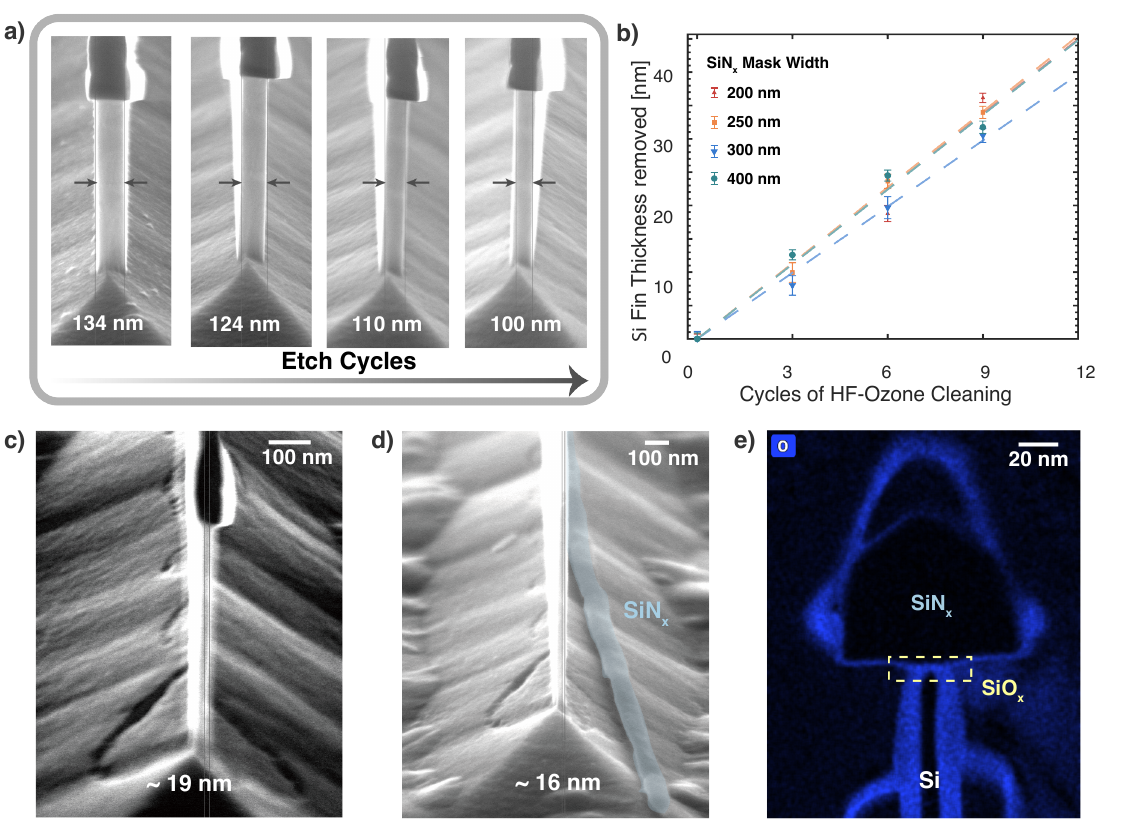}
\caption{\label{fig:2} (a) Cross-sectional SEM images of Si fins after KOH, 3, 6, and 9 cycles of UV ozone and then HF etching, showing progressive reduction in the fin thickness for a SiN$_x$ mask width of 250~nm. (b) Fin thickness removed as a function of etching cycles for various SiN$_x$ mask widths (400, 300, 250, and 200~nm), all etched in KOH for 1.5~min, demonstrating similar etch rates for various SiN$_x$ mask widths. (c) Cross-sectional SEM of a $\sim$~19~nm  Si fin. (d) Cross-sectional SEM of a $\sim$~16~nm Si fin showing SiN$_x$ mask detachment. (e) Cross-sectional STEM-EDS mapping of the Si/SiN$_x$ interface of an RTA-VHF-processed fin in which the final oxide was intentionally not fully removed, revealing a thin interfacial SiO$_x$ layer. Note: For each data point, the fin thickness was measured at four positions near the mid-height of the same cross-section. The plotted values represent the mean of the four measurements, and the error bars represent one standard deviation of the four local thickness measurements.}

\end{figure*}

The fabrication of the sub-10~nm Si fins includes two primary steps: fin patterning and wet chemical etching followed by dry gaseous digital etching. The initial patterning and wet etching form a fin-like structure, and the following dry etching step controllably reduces the fin thickness to below 10~nm. The process starts with a 3-inch float-zone Si (110) substrate with a 100~nm low-stress low-pressure chemical vapor deposition (LPCVD) deposited silicon nitride layer (SiN$_x$) on top. With this orientation, there are vertical $\{111\}$ planes parallel to the wafer flat and perpendicular to the (110) surface, which is critical for the anisotropic KOH etch discussed later \cite{seidel1990anisotropic}. The SiN$_x$ not only serves as an etch mask, but also functions as a shadow mask to form a break in the superconducting wiring layers during the superconductor angle deposition to form JJs. In the absence of the SiN$_x$ mask, the superconductor deposited afterward would bridge across the Si fin, shorting the capacitor or junction. Thus, the structural integrity of the SiN$_x$ throughout the fabrication process is crucial. Alignment marks aligned to the wafer flat ($\langle 111\rangle$) are patterned on the wafer with photoresist using the automatic alignment feature of a maskless aligner. Gold alignment marks are formed on the SiN$_x$ layer by electron beam evaporation and lift-off of the photoresist. Electron beam lithography and dry etching are utilized to define patterns on the SiN$_x$ hard mask. Initially, the pattern dimensions were rectangles with widths ranging from 100~nm to 400~nm (in 50~nm steps) and lengths of 200~$\upmu$m. The pattern later changed to rectangles with widths from 125~nm to 210~nm (in 5~nm steps) and lengths of 200~$\upmu$m. By varying the width, the thinnest fins are intentionally designed to collapse upon etching, while the next wider fin is then the optimal dimension at which the fin remains intact while being the thinnest. The hard mask is then etched in an inductively coupled plasma (ICP) etcher using CF$_4$/O$_2$ gases. The 3-inch wafer is then coated with photoresist and diced into multiple 1 $\mathrm{cm^2}$ pieces for process development. After dicing, the protective photoresist is stripped by ultrasonication in N-methyl-2-pyrrolidone at 80 $^{\circ}$C, followed by a solvent clean in acetone and isopropanol. Before KOH etching, each sample is first treated with a 5~min O$_2$ plasma, followed by a 10~s plasma etch using a C$_4$F$_8$/SF$_6$/CF$_4$ gas mixture, and finished with another 5~min O$_2$ plasma in an ICP etcher to prevent roughness in the Si (110) surface after KOH etching.

After cleaning, the Si is wet etched in potassium hydroxide solution (KOH, 45 wt\% in H$_2$O, 87$^{\circ}$C) for 30~s to 2~min to form the fin-like structures with smooth $\{111\}$ sidewalls. Si fins shown in Figure 2(a) and (b) are etched in KOH for 1.5~min, Si fins in Figure 2(c) and (d) are etched in KOH for 30~s, and the fins in Figures 3, 4, and 5 are etched in KOH for 2~min. 1.5~min of anisotropic KOH wet etching of Si fins patterned with a 200~nm wide SiN$_x$ mask typically yields a fin thickness of approximately 74~nm with some tapering after 1~$\upmu$m of vertical etching.

\begin{figure*}[t]
\includegraphics[width=0.9\textwidth]{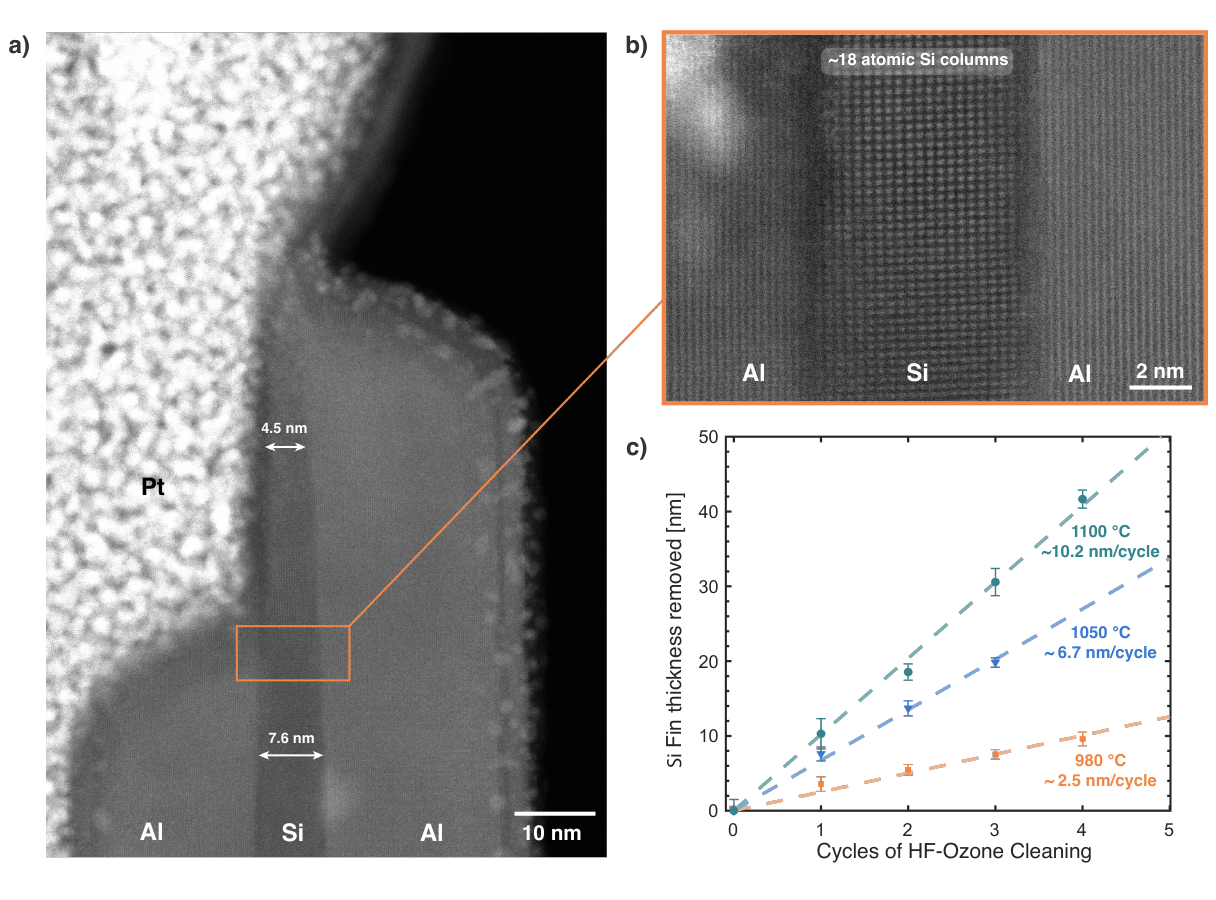}
 \caption{\label{fig:3} (a) Cross-sectional STEM image of a high-aspect-ratio Si fin after RTA-VHF thinning, demonstrating a thickness of $\sim$~4.5~nm near the top and $\sim$~6~nm near the start of the junction. (b) Zoomed-in area of (a) with Si barriers of $\sim$~18 atomic columns ($\sim$~6~nm). (c) Thickness removed of Si fins as a function of RTA-VHF cycles performed at various annealing temperatures, illustrating the tunable etch rate for the process. Note: For each data point, the fin thickness was measured at four positions near the mid-height of the same cross-section. The plotted values represent the mean of the four measurements, and the error bars represent one standard deviation of the four local thickness measurements.}
\end{figure*}

An effective method to further reduce the fin thickness is a digital etching process with repetitive room-temperature UV ozone exposure and liquid HF etching, which is highly selective in etching SiO$_x$ over Si. 10 minutes of UV ozone was utilized to oxidize the Si fin, and 60~s of a 4.9\% hydrofluoric acid (HF) etch was performed to remove the silicon oxide (SiO$_x$, x$\sim$~2). During each cycle, the UV ozone exposure forms a SiO$_x$ layer on the Si(111) surface, which was subsequently etched by HF, and the oxidation and etching cycle was repeated until the desired thickness was achieved. The thickness of the Si fins was verified using scanning electron microscopy (SEM) and scanning transmission electron microscopy (STEM). Notably, for fins below 10~nm, edge effects and charging of the high-aspect-ratio structure prevent the SEM from providing reliable thickness measurements. Therefore, cross-sectional STEM is utilized to quantify the fin dimensions. Prior to FIB preparation, Al was deposited onto both fin sidewalls at an incidence angle of $\pm$25~$^\circ$ relative to the vertical direction, producing a geometry compatible with future JJ fabrication.

Figure 2(a) shows SEM images taken along the $\langle 1\,\overline{1}\,0 \rangle$ direction indicating the progressive reduction of the fin thickness after KOH, 3, 6, and 9 etching cycles, respectively. The fin reduction thickness for various pattern widths is plotted as a function of the number of UV Ozone-HF cycles, and the fin thickness thinning rates for different widths are consistently around 3 $\sim$~4 nm/cycle (Figure 2(b)), confirming a similar etch rate for different initial SiN$_x$ mask widths. 

In principle, this digital etching process enables the fin thickness to be reduced to any desired thickness. In practice, however, the process is limited by SiN$_x$ hard mask detachment at around $\sim$~19~nm. Additional etching cycles beyond this range lead to SiN$_x$ detaching from the fin. Figure 2(c) shows a Si fin with a thickness of approximately 19~nm after five cycles of 10 min UV-ozone oxidation and 60~s 4.9\% HF etch. After one additional etching cycle, the fin thickness is further reduced to approximately 16~nm. However, the SiN$_x$ hard mask detaches, as highlighted in blue in Figure 2(d). 

To understand the origin of the SiN$_x$ mask detachment, cross-sectional STEM with energy-dispersive X-ray spectroscopy was performed on a Si fin thinned by digital etching using RTA oxidation and VHF etching (Figure 2(e)). To investigate the presence of an interfacial oxide, the VHF step was intentionally shortened so that the oxide was not fully removed. Figure 2(e) shows a thin SiO$_x$ later between the LPCVD SiN$_x$ mask and the Si fin. Although this Si fin sample was prepared by RTA oxidation and VHF etching, a similar SiO$_x$ interfacial layer is also expected in the UV-ozone/HF Si fins, since the SiO$_x$ layer is likely to originate either from the native oxide before the LPCVD SiN$_x$ deposition or from lateral oxidation of the Si fin during the oxidation cycle. The high SiO$_x$ etch rate in liquid HF compared to VHF results in lateral undercutting of this interfacial oxide more rapidly, contributing to SiN$_x$ mask detachment. In addition, dipping the sample multiple times in liquid HF and rinsing in DI water could inflict mechanical damage on the already-fragile structure, increasing the possibility of SiN$_x$ hard mask detachment.

To further thin the fins to the sub-10~nm regime while maintaining the structural integrity of the SiN$_x$ requires reducing the exposure of the interfacial oxide to the etchant. RTA oxidation and VHF were implemented to achieve these goals. RTA oxidation provides a wider range of tunability by adjusting the process temperature, while VHF etches SiO$_x$ at a much slower rate than liquid HF, thus reducing undercutting of the interfacial SiO$_x$ between the SiN$_x$ mask and the top of the Si fin. The SiO$_x$ etch rates are 20~nm/min for VHF and 80~nm/min for 4.9\% HF, and both are calibrated on RTA-grown oxide on a planar Si(110) sample. Figure 3(c) demonstrates the temperature dependence of Si consumption during the RTA oxidation, and the Si thinning can be tuned by adjusting the annealing temperature. The process was calibrated by ramping the samples in an O$_2$ gas and oxidizing at 1100~$^{\circ}$C, 1050~$^{\circ}$C, and 980~$^{\circ}$C for 1 min in pure O$_2$ gas \cite{mohadjeri1998oxidation}, followed by 80~s of VHF etching. The corresponding fin thinning rates are $\sim$~10.2~nm/cycle, 6.7~nm/cycle, and 2.5~nm/cycle. High-temperature oxidation provides rapid thinning when fins are more than 50~nm, while lower-temperature oxidation enables fine control as the sub-10~nm regime is approached. By combining these two regimes, a high-aspect-ratio Si fin with a thickness of 6~nm was achieved, as confirmed by cross-sectional STEM (Figure 3(a)(b)).

\begin{figure*}[t]
\includegraphics[width=0.9\textwidth]{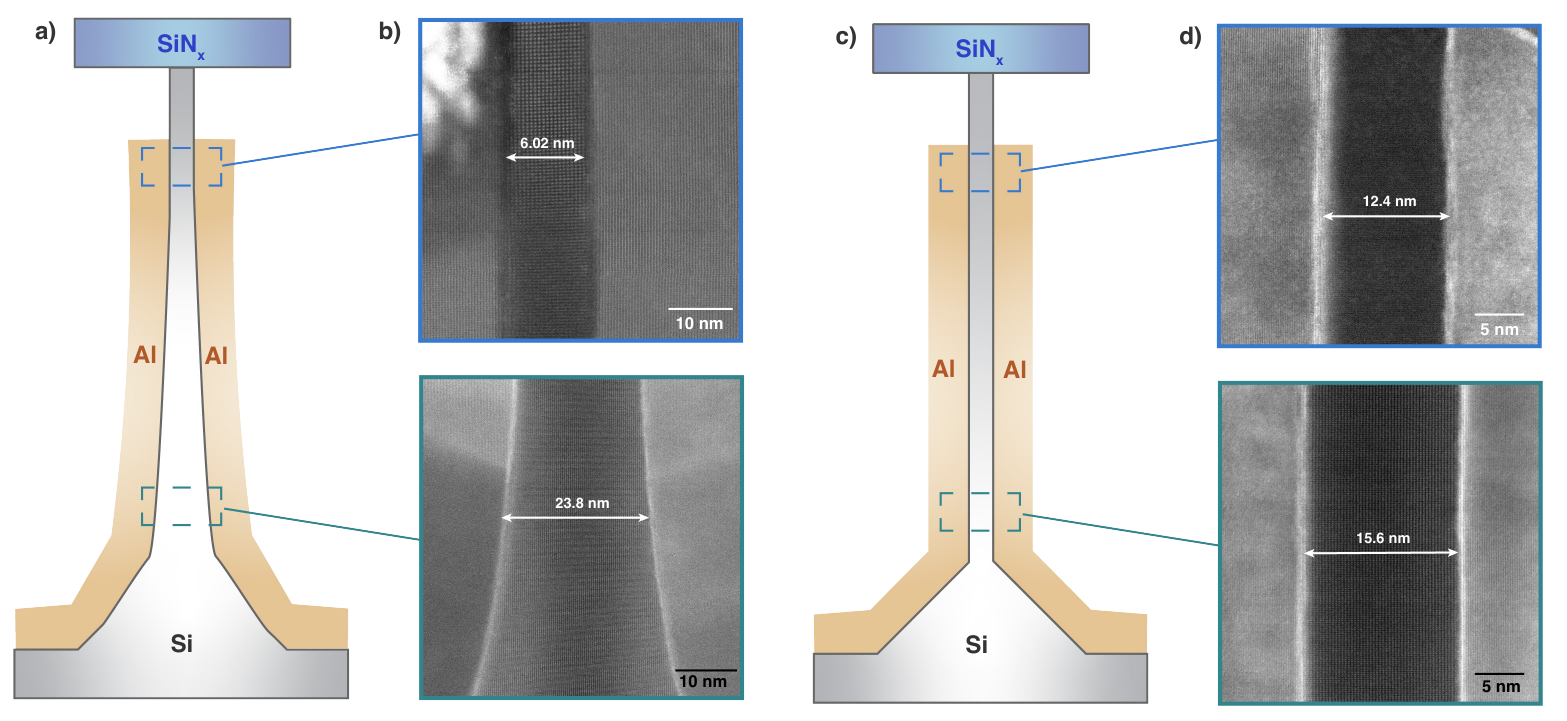}
\caption{\label{fig:4} (a) Schematic of the Si fin cross-section demonstrating tapering. (b) Cross-sectional STEM image of the Si barrier at the top and near the substrate showing a difference of $\sim$~17~nm with a ramping condition that uses pure O$_2$ at 10~$^{\circ}$C/s (c) Schematic of the Si fin cross-section demonstrating uniformity. (d) Cross-sectional STEM image of the Si barrier at the top (12.4~nm) and near the substrate (15.6~nm) showing a difference of only $\sim$~3 nm with ramping conditions using N$_2$/O$_2$ (5:1) at 2~$^{\circ}$C/s. Both fins were thinned using the same RTA oxidation sequence starting with three cycles of 1 min oxidation at 1100~$^{\circ}$C followed by one cycle of 1 min oxidation at 1050~$^{\circ}$C and one cycle of 1 min oxidation at 980~$^{\circ}$C, with VHF etch times of 80~s per cycle for (b) and 120~s per cycle for (d). }
\end{figure*}

Figure 4(a) shows a schematic, and Figure 4(b) shows high-resolution STEM images of the top and bottom portions of a fin thinned using a pure O$_2$ ramp at 10~$^{\circ}$C/s. The top portion of the fin is reduced to $\sim$~6~nm, whereas the bottom near the substrate remains substantially thicker at $\sim$~23~nm. This $\sim$~17~nm thickness gradient indicates that only the top portion of the fin would dominate the tunneling current and the capacitance of the fin. Tapering was also observed in STEM images after KOH etching alone due to sidewall regions near the top of the fin being etched for longer than those formed later near the base \cite{seidel1990anisotropic}. 

It was speculated that the large temperature gradient during the RTA ramping process could result in premature nonuniform oxidation along the fin. Hence, to mitigate these effects, both the temperature ramp rate and the ambient gas composition were changed during the RTA oxidation process. First, the ramp rate was reduced from 10~$^{\circ}$C/s to 2~$^{\circ}$C/s to reduce the temperature gradients. Second, the ambient gas during the ramp was changed from pure O$_2$ to N$_2$/O$_2$ (5:1), lowering the O$_2$ partial pressure and thereby limiting oxidation before the setpoint temperature was reached and the gas changed to pure O$_2$.

\begin{figure*}[t]
\centering
\includegraphics[width=0.9\textwidth]{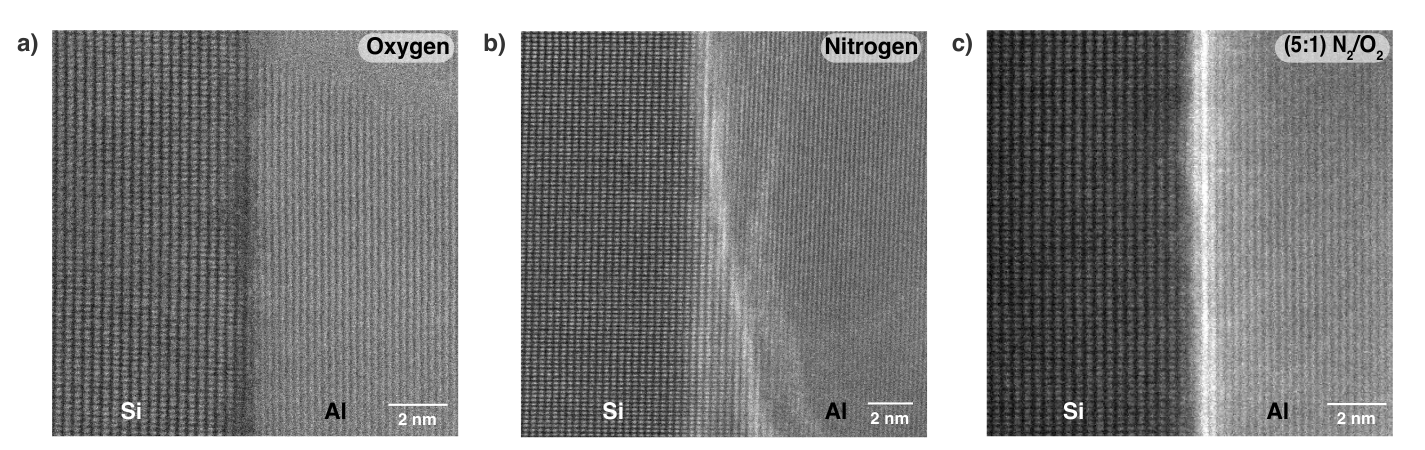}
 \caption{\label{fig:5} Cross-sectional STEM image of the Si/Al interface processed by rapid thermal annealing ramping with (a) oxygen, (b) nitrogen, and (c) nitrogen/oxygen (5:1). Note: Oxygen was not observed at the Al/Si interface by EDS. However, a slight misalignment in the zone axis can lead to a change in the contrast at the interface.}
\end{figure*}

Initially, pure N$_2$ was used during the ramping process. However, this led to additional surface roughness and inhibition of oxide etching during the VHF process (Figure 5(b)). In a pure N$_2$ environment, the low O$_{2}$ partial pressure could lead to desorption of SiO, resulting Si etching and poor surface morphology \cite{mohadjeri1998oxidation} instead of SiO$_{x}$ formation. 

In addition to SiO desorption, annealing in a pure N$_2$ ambient environment could lead to nitridation of the Si surface. It was speculated that the residual oxygen would mediate the Si-N reaction, resulting in the formation of SiO$_x$N$_y$ \cite{green1997ultrathin}. Nitrogen incorporation strongly reduces the etch rate of SiO$_x$N$_y$ in a hydrofluoric buffer oxide etchant, approaching zero at high nitrogen content \cite{topka2022critical}, and VHF etches SiN$_x$ far more slowly than SiO$_x$ \cite{shimaoka2006characteristics}. Therefore, formation of a nitrogen-containing SiO$_x$N$_y$ layer during the ramp could inhibit removal of the SiO$_x$ during the subsequent VHF step.

To limit premature oxidation during the ramp while suppressing SiO desorption and nitrogen incorporation, a mixed nitrogen-oxygen ambient (5:1) during the ramp was introduced. Under this condition, the O$_2$ partial pressure remains well above the critical pressure for oxide growth, forming a SiO$_x$ film rather than forming and desorbing volatile SiO that would etch the Si \cite{smith1982reaction}. Adsorbed oxygen nucleates SiO$_x$ islands that subsequently grow and merge into a uniform oxide film \cite{smith1982reaction}. A protective SiO$_x$ layer is formed, resulting in a smooth SiO$_x$/Si interface \cite{mohadjeri1998oxidation}, minimizing surface roughness after VHF removal. As shown in Figures 5(a) and 5(c), the cross-sectional STEM indicates qualitatively similar interface morphology for the O$_2$ and N$_2$/O$_2$ ramp conditions, whereas the pure N$_2$ condition exhibits visibly greater surface roughness.

These adjustments resulted in fins with improved uniformity, as demonstrated by cross-sectional STEM (Figure 4(d)). The tapering profile was reduced from $\sim$~17~nm to $\sim$~3~nm on a 1.2~$\upmu$m tall fin measuring $\sim$~12~nm near the top and $\sim$~15~nm near the base. Both fins in Figure 4 were thinned utilizing the same oxidation sequence starting with three cycles of 1~min oxidation at 1100~$^{\circ}$C followed by one cycle of 1~min oxidation at 1050~$^{\circ}$C and one cycle of 1~min oxidation at 980~$^{\circ}$C. The VHF step was 80~s per cycle for the fin in Figure 4(b) and 120~s per cycle for the fin in Figure 4(d). A planar Si(110) witness sample processed alongside each fin was hydrophobic after every VHF step, indicating complete removal of the thermal oxide. Since VHF is highly selective to SiO$_x$ over Si, the additional 40~s is not expected to consume additional Si once the oxide is removed, and the tapering profile is therefore attributed to the ramp conditions. Thus, these conditions and further optimization of the O$_2$ fraction with RTA oxidation–VHF etching has the potential to produce uniform fins in the sub-10~nm regime, enabling more consistent junction areas.

In this work, a digital etching methodology was developed for thinning high-aspect-ratio single-crystalline Si fins down to the sub-10~nm regime. UV-ozone oxidation followed by liquid HF oxide removal enabled thinning to approximately 20~nm before SiN$_x$ detachment. However, the fin retained the tapering profile that resulted from the anisotropic KOH wet etching. Under a fixed RTA oxidation sequence, changing the RTA ramp condition from pure O$_2$ at 10~$^{\circ}$C/s to N$_2$/O$_2$ at 2~$^{\circ}$C/s reduced the tapering from approximately 17~nm to approximately 3~nm on a fin with a thickness of approximately 12~nm near the top and 15~nm near the base. These results, combined with further optimization of the O$_2$ fraction to enable more consistent junction uniformity, establish a pathway toward achieving uniformly sub-10~nm single-crystalline Si fins for Josephson-junction applications.

Further improvement in the uniformity can be achieved by tuning the O$_2$ fraction during the ramp, balancing suppression of premature oxidation against the critical pressure to prevent SiO desorption. An additional atomic layer etching (ALE) step at the end of the process could be implemented to further improve surface smoothness through self-limiting, layer-by-layer removal \cite{abdulagatov2018thermal}. A final etch in buffered NH$_4$F could additionally be used to achieve atomically flat, hydrogen-terminated Si(111) sidewalls \cite{higashi1990ideal,higashi1991comparison}. Overall, the fabrication process demonstrated here is a crucial step toward realizing vertical JJs using single-crystalline Si fins.

\begin{acknowledgments}
We acknowledge the support of the New and Emerging Qubit Science and Technology (NEQST) Program initiated by the U.S. Army Research Office (ARO) with Grant No. W911NF2210052, National Science Foundation (NSF) Quantum Foundry through Q-AMASE-i Program via Award No. DMR-1906325, and the National Institute of Standards and Technology (NIST).
We also acknowledge the use of the shared facilities of the UCSB Materials Research Science and Engineering Centers (No. NSF DMR–2308708) and the Nanotech UCSB Nanofabrication facility. We would like to thank David Pappas and Aranya Goswami for their useful discussion as well as James Beall for assistance with LPCVD silicon nitride growth. 
\end{acknowledgments}

\section*{Data Availability Statement}
The data that support the findings of this study are available from the corresponding author upon reasonable request.

\section*{References}
\bibliography{aipsamp}

\end{document}